\documentclass[11pt]{article}
\usepackage{fullpage,latexsym,amsmath,amssymb}

\newcommand{\EPR}{{\rm EPR}}

\newcommand{\dirprod}{\otimes}
\newcommand{\reals}{{\mathbb R}}

\newcommand{\complexes}{{\mathbb C}}
\newcommand{\cH}{{\cal H}}

\newcommand{\trace}{{\rm tr}}
\newcommand{\map}[3]{{#1}:{#2}\rightarrow{#3}}
\newcommand{\bra}[1]{\left\langle{#1}\right|}
\newcommand{\ket}[1]{\left|{#1}\right\rangle}

\renewcommand{\bowtie}[2]{\ket{#1}\!\bra{#2}}
\newcommand{\lbt}[2]{{\ket{#1}_{#2}\!\bra{#1}}}

\newcommand{\prop}{\propto}
\newcommand{\setof}[1]{{\left\{{#1}\right\}}}
\newcommand{\eqdf}{\mathrel{=_{\rm df}}}
\newcommand{\two}{\setof{0,1}}
\newcommand{\four}{\setof{0,1,2,3}}
\newcommand{\adj}[1]{{{#1}^{\dagger}}}
\newcommand{\septwo}[4]{{\oplus\left[\left({#1},{#2}\right),\;%
\left({#3},{#4}\right)\right]}}
\newcommand{\septhree}[6]{{\oplus\left[\left({#1},{#2}\right),\;%
\left({#3},{#4}\right),\;\left({#5},{#6}\right)\right]}}

\newtheorem{definition}{Definition}

\newtheorem{theorem}[definition]{Theorem}

\title{Universal quantum computation with two- and three-qubit
projective measurements}

\author{Stephen A. Fenner\thanks{Computer Science and Engineering
Department, Columbia, SC  29208.  Email: fenner@cse.sc.edu.  Research
partially supported by South Carolina Commission on Higher Education
Research Initiation Grant R-01-0256.}\\
University of South Carolina
\and
Yong Zhang\thanks{Computer Science and Engineering Department,
Columbia, SC  29208.  Email: zhang29@cse.sc.edu.  Partially supported
by South Carolina CHE SCRIG Grant R-01-0256.}\\
University of South Carolina
}

\date{\today}

\begin{document}

\maketitle

\begin{abstract}

We present a finite set of projective measurements that, together with
quantum memory and preparation of the $\ket{0}$ state, suffice for
universal quantum computation.  This extends work of Nielsen
\cite{Nielsen:measurements}, who proposed a scheme in which an
arbitrary unitary operation on $n$ qubits can be simulated using only
projective measurements on at most $2n$ qubits.  All measurements in
our set involve two qubits, except two measurements which involve
three qubits.  Thus we improve by one the upper bound, implied by
Nielsen's results, on the maximum number of qubits needed to
participate in any single measurement to achieve universal quantum
computation.

Each of our measurements is two-valued, and each can be expressed
mathematically as a Boolean combination of single-qubit measurements.
\end{abstract}

\section{Introduction}

A major goal of quantum information processing is to find a minimal
set of primitive quantum operations that are simple enough to be
implemented easily and yet are universal for quantum computation.  It
has long been known that quantum circuits employing gates drawn from
small families of unitary operators can efficiently simulate any
``reasonable'' quantum computation (on a time-bounded quantum Turing
machine, say) to arbitrarily close approximation
\cite{Yao:q-circuits,BBCDMSSSW:quantum-gates,BMPR:fault-tolerant}.

Nielsen recently proposed a scheme for universal quantum computation
where no unitary operators are used at all.  Instead, an arbitrary
unitary quantum gate on $n$ qubits is simulated by a protocol
involving only projective measurements on $2n$ qubits, together with
quantum memory, preparation of the $\ket{0}$ state, and classical
communication \cite{Nielsen:measurements}, thus showing that projective
measurements on at most four qubits suffice for universal quantum
computation.  We describe his scheme in Section~\ref{sec:Nielsen}.

We build on Nielsen's idea to show that only a finite set of
projective measurements---each on at most three qubits---are needed
for universal quantum computation.  We isolate a finite set $S$ of
projective measurements such that
\begin{itemize}
\item
$S$ is universal for quantum computation (in Nielsen's framework),
\item
all measurements in $S$ involve at most two qubits, except two which
involve three qubits,
\item
each measurement of $S$ is \emph{binary}, that is, has two possible
outcomes ($0$ and $1$, say) with the same amount of degeneracy for
each value, and
\item
each measurement in $S$ can be expressed as a Boolean combination of
results from single-qubit projective measurements drawn from a set of
cardinality four.
\end{itemize}
(See Section~\ref{sec:summary} for a summary of the measurements
used.)  As a corollary, three-qubit measurements suffice for universal
quantum computation.

Our results follow from analyzing Nielsen's scheme.  This scheme
simulates an arbitrary $n$-qubit unitary operation $U$ (for
$n\in\{1,2\}$) via a series of complete (nondegenerate) projective
measurements on $2n$ qubits.  We decompose each of these measurements
into a sequence of $2n$ pairwise commuting binary
measurements---extracting one classical bit per measurement.  This
decomposition allows some leeway over which projections to combine for
each binary measurement.  By choosing the right combinations, we can
express our binary measurements in a particularly elegant form:
single-qubit measurement results combined with a Boolean operator.  We
call such measurements \emph{pseudoseparate}.  We are able to perform
this decomposition into pseudoseparate measurements to simulate
arbitrary one-qubit unitary operations and the two-qubit controlled
NOT (C-NOT) gate.  It is with C-NOT that we find that measurements on
four qubits are not needed.

\section{Preliminaries}

We assume knowledge of the basic concepts and notation used in quantum
computation, as found in, for example, Nielsen and Chuang
\cite{NC:quantumbook}.

If $A$ and $B$ are either both vectors or both operators, then we
write $A\prop B$ to mean that $A$ and $B$ are equal up to a phase
factor: $A = e^{i\theta} B$ for some $\theta\in\reals$.

Let $\sigma_0 = I$, $\sigma_1 = \sigma_x$, $\sigma_2 = \sigma_y$, and
$\sigma_3 = \sigma_z$ be the usual one-qubit Pauli spin operators.
Following standard practice, if $\vec{n} = (n_1,n_2,n_3)$ is a vector
in $\reals^3$, we let $\vec{n}\cdot\vec{\sigma}$ denote $n_1\sigma_1 +
n_2\sigma_2 + n_3\sigma_3$.  For $i,j\in\four$, we define
$[i,j]\in\four$ to be so that $\sigma_i\sigma_j \prop \sigma_{[i,j]}$.
That is, $[i,j] = [j,i]$, \ $[0,j] = j$, \ $[i,i] = 0$, and if
$i,j\in\setof{1,2,3}$ and $i\neq j$, then $[i,j]$ is the unique
element of $\setof{1,2,3} - \setof{i,j}$.

We define
\begin{equation}
\ket{\EPR} = \frac{1}{\sqrt{2}} (\ket{00} + \ket{11}),
\end{equation}
and for $i\in\four$ define the Bell states
\begin{equation}
\ket{B_i} = (I\dirprod\sigma_i)\ket{\EPR}.
\end{equation}

We will be informal and use $I$ to denote the identity operator on any
Hilbert space, sometimes two different spaces in the same equation.
Which identity operator is intended should be clear from the context.

Qubits will often be labeled with numbers $1,2,3,\ldots$, and we will
sometimes put numerical subscripts on quantum states (and operators)
to show which qubits they refer to (or act upon).  For example, the
state $\ket{B_0}_{13}\ket{B_3}_{24}$ can be written out as
\[ \frac{1}{2}(\ket{0000} - \ket{0101} + \ket{1010} - \ket{1111}), \]
and the operator $U_{12} \dirprod V_{34}$ means ``apply $U$ to qubits 1
and 2, and $V$ to qubits 3 and 4.''

Let $\map{f}{\two^n}{\two}$ be an $n$-ary Boolean
function for $n \geq 1$.  We say that $f$ is \emph{balanced} if
$f^{-1}(0)$ and $f^{-1}(1)$ both have cardinality $2^{n-1}$.

\subsection{Projective Measurements}

A projective measurement on a Hilbert space $\cH$ of $n$ qubits
corresponds to a complete $k$-tuple $(P_0,\ldots,P_{k-1})$ of mutually
annihilating projection operators on $\cH$, that is, $P_iP_j =
\delta_{ij}P_i$ and $\sum_i P_i = I$.  The projector $P_i$ corresponds
to getting the classical result $i$.  If the system is in state $\rho$
and the measurement is made, we will see result $i$ with probability
$p_i = \trace(P_i\rho)$ and in such a case, the state collapses to
$P_i\rho P_i/p_i$.  We will say that the measurement is \emph{binary}
if $k=2$ and $\trace P_0 = \trace P_1 = 2^{n-1}$.

A single-qubit projector with unit trace can always be expressed in
the form $(I + \vec{\alpha}\cdot\vec{\sigma})/2$ for some unit vector
$\vec{\alpha} \in \reals^3$.

Clearly, any nondegenerate projective measurement on $n$ qubits---that
is, one where $k=2^n$ and $\trace P_i = 1$ for all $i$---is
\emph{equivalent} to some sequence of $n$ pairwise commuting binary
projective measurements, in that they yield the same classical
information and resulting distribution of quantum states.  For
example, a nondegenerate measurement $(P_0,P_1,P_2,P_3)$ on two qubits
is equivalent to first measuring by $(P_0+P_1,P_2+P_3)$ then by
$(P_0+P_2,P_1+P_3)$.

\begin{definition}\label{def:f-separate}
Let $\map{f}{\two^n}{\two}$ be a balanced Boolean function.  A binary
projective measurement $(P_0,P_1)$ on $n$ qubits is
\emph{$f$-separate} if there exist $n$ single-qubit binary projective
measurements $(P_0^{(1)},P_1^{(1)}),\ldots,(P_0^{(n)},P_1^{(n)})$ such
that for $i\in\two$,
\[ P_i = \sum_{j_1,\ldots,j_n\in\two\colon f(j_1\cdots j_n) = i}
P_{j_1}^{(1)} \dirprod \cdots \dirprod P_{j_n}^{(n)}. \]
If this is the case, we denote the measurement $(P_0,P_1)$ as
$f\left[(P_0^{(1)},P_1^{(1)}),\ldots,(P_0^{(n)},P_1^{(n)})\right]$.  A
binary projective measurement is \emph{pseudoseparate} if it is
$f$-separate for some balanced $f$.
\end{definition}

All but one of the measurements in our universal set are $f$-separate,
where $f$ is the parity function $\oplus$ (exclusive OR).  We can view
an $f$-separate measurement intuitively as single-qubit measurements
combined in a classical way.  For example, suppose $n=2$ and we have
two qubits $A$ and $B$ belonging to Alice and Bob, respectively.
Alice performs some projective measurement $(P_0^A,P_1^A)$ on her
qubit, getting the ``classical'' bit $j_A$, and Bob independently
measures his qubit according to some $(P_0^B,P_1^B)$, getting $j_B$.
Alice and Bob communicate $j_A$ and $j_B$ to a third party, Carol, who
computes and outputs the classical bit $f(j_Aj_B)$.  Thus the
classical result of the whole measurement is $f(j_Aj_B)$, and the
resulting quantum state is the projection onto the subspace consistent
with this classical result.

It should be stressed that pseudoseparate measurements are not the
same as truly separate single-qubit measurements.  The bits $j_A$ and
$j_B$ communicated by Alice and Bob are not really classical---Alice,
Bob, and Carol are essentially quantum agents who must work in
isolation from the environment.  If Alice and Bob shared their bits
with the (macroscopic) environment, as is the case with truly
classical bits, then the degeneracy of the measurement would be lost,
and we'd get two classical bits as a result of two completely separate
measurements.  This would not do for universal quantum computation,
which needs to create entanglement between $A$ and $B$.  It is not
clear at this point whether pseudoseparate measurements are any easier
to implement than other projective measurements, but their
mathematical simplicity is attractive nonetheless, and gives some hope
for an easier implementation.

For $n\geq 2$, almost all $n$-qubit binary projective measurements are
\emph{not} pseudoseparate.  This can be seen by counting the number of
continuous degrees of freedom for the two respective measurement
types.  The number of complex degrees of freedom for an $n$-qubit
binary projective measurement is $2^{2(n-1)}$, the dimension of the
Grassmann manifold $G_{2^{n-1},2^n}(\complexes)$
\cite{Fujii:Grassmann}, whereas the number of continuous degrees of
freedom for an $n$-qubit pseudoseparate measurement is only $n$, i.e.,
one for each single-qubit measurement (the choice of $f$ is discrete
and does not add to the continuous degrees of freedom).

\subsection{Nielsen's Scheme}\label{sec:Nielsen}

Here we briefly review Nielsen's protocol for simulating an arbitrary
$n$-qubit unitary gate $U$ by projective measurements, for
$n\in\setof{1,2}$ \cite{Nielsen:measurements}.  The scheme is a
generalization of simple quantum teleportation
\cite{BBCJPW:teleportation}.

We first consider the case for $n=1$.  We are given a single qubit
state $\ket{\psi}$ and we wish to produce $U\ket{\psi}$, at least up
to a phase factor.  First we prepare two ancilla qubits off line in
one of the four states
\begin{equation}\label{Uj}
\ket{U_j} = (I \dirprod U\sigma_j)\ket{\EPR} = (I \dirprod
U)\ket{B_j}
\end{equation}
(for some $j\in\four$) by measuring in this basis.  We then perform a
Bell measurement (basis
$\setof{\ket{B_0},\ket{B_1},\ket{B_2},\ket{B_3}}$) on the combined
system of $\ket{\psi}$ and
the first of the ancilla qubits, giving a classical result $m\in\four$
corresponding to $\ket{B_m}$.  Each $m$ occurs with probability $1/4$
independent of $j$, and the resulting state of the second ancilla is
then $U\sigma_j\sigma_m\ket{\psi} \prop U\sigma_{[j,m]}\ket{\psi}$.
With probability $1/4$, we have $m=j$ and so we have succeeded in
producing $U\ket{\psi}$ in the second ancilla.  If not, then we repeat
the protocol, this time with input $U\sigma_j\sigma_m\ket{\psi}$,
attempting to simulate the gate
\begin{equation}\label{next-U}
U' = U\sigma_m\sigma_j\adj{U}.
\end{equation}
(We can't start over with $\ket{\psi}$, since this state may be
difficult to produce in quantity.)  Again, we will produce
$U\ket{\psi}$ with probability $1/4$, but if not, we continue to
repeat the process, each time trying to undo the error of the last
trial.  Thus the expected number of trials before success is four.

For the case $n=2$, we are given a two-qubit input state $\ket{\psi}$
and wish to simulate a two-qubit unitary gate $U$.  By a suitable
projective measurement, we prepare four ancilla qubits off line in one
of the sixteen states
\begin{equation}\label{Ujk}
\ket{U_{jk}} = (I_{12} \dirprod (U(\sigma_j \dirprod
\sigma_k))_{34})\ket{\EPR}_{13}\ket{\EPR}_{24} = (I_{12} \dirprod
U_{34})\ket{B_j}_{13}\ket{B_k}_{24}
\end{equation}
for some $j,k\in\four$.  Relabel the qubits so that $\ket{\psi}$ is on
qubits $1$ and $2$, and the first two ancilla qubits are $3$ and $4$.
We now do two separate Bell measurements, the first on qubits $1$ and
$3$ giving the classical result $m\in\four$, and the second on qubits
$2$ and $4$ giving the classical outcome $n\in\four$.  (Each
combination $(m,n)$ occurs with probability $1/16$.)  The resulting
state of the last two ancilla qubits will then be $\ket{\psi'} =
U(\sigma_j\sigma_m \dirprod \sigma_k\sigma_n)\ket{\psi}$.  If
$(j,k)\neq (m,n)$---which occurs with probability $15/16$---then the
protocol is repeated with input state $\ket{\psi'}$, simulating
\[ U' = U(\sigma_m\sigma_j \dirprod \sigma_n\sigma_k)\adj{U}, \]
and so on.

For any $\epsilon > 0$, we need ${\cal O}(\log\frac{1}{\epsilon})$
trials to get a failure rate below $\epsilon$.

\section{Main Results}\label{sec:main}

We will now build our universal family of binary projective
measurements.  We consider the finite universal family of gates
containing only the C-NOT gate, the one-qubit Hadamard gate
\[ H = \frac{1}{\sqrt{2}}
\left[\begin{array}{rr} 1 & 1 \\ 1 & -1 \end{array}\right], \]
and the one-qubit $\pi/8$ gate
\[ T = e^{-i\pi\sigma_3/8} \prop \left[\begin{matrix} 1 & 0 \\ 0 &
e^{i\pi/4} \end{matrix}\right] \]
\cite{BMPR:fault-tolerant,NC:quantumbook}.  We
only need to show that these gates can be simulated using a finite set
of projective measurements.  We first describe the case for
single-qubit gates.

\subsection{Simulating One-Qubit Gates}

In order to run a single trial of Nielsen's protocol to simulate a
one-qubit gate $U$, we need a Bell measurement and the complete binary
measurement in the two-qubit basis states $\ket{U_j}$ of
Equation~\ref{Uj}, which corresponds to projectors
$(\bowtie{U_0}{U_0}, \bowtie{U_1}{U_1}, \bowtie{U_2}{U_2},
\bowtie{U_3}{U_3})$.  Note that the Bell measurement itself is just a
special case of Equation~\ref{Uj} where $U = I$.  We consider this
special case first, from which the general case can easily be derived.

Noting that
\[ \bowtie{\EPR}{\EPR} = \frac{1}{4}(I\dirprod I + \sigma_1\dirprod
\sigma_1 - \sigma_2 \dirprod \sigma_2 +\sigma_3 \dirprod \sigma_3), \]
it is routine to calculate each $\bowtie{B_j}{B_j} = (I \dirprod
\sigma_j)\bowtie{\EPR}{\EPR}(I \dirprod \sigma_j)$:
\begin{eqnarray*}
\bowtie{B_0}{B_0} & = & (I + \sigma_1\dirprod\sigma_1 -
\sigma_2\dirprod\sigma_2 + \sigma_3\dirprod\sigma_3)/4 \\
\bowtie{B_1}{B_1} & = & (I + \sigma_1\dirprod\sigma_1 +
\sigma_2\dirprod\sigma_2 - \sigma_3\dirprod\sigma_3)/4 \\
\bowtie{B_2}{B_2} & = & (I - \sigma_1\dirprod\sigma_1 -
\sigma_2\dirprod\sigma_2 - \sigma_3\dirprod\sigma_3)/4 \\
\bowtie{B_3}{B_3} & = & (I - \sigma_1\dirprod\sigma_1 +
\sigma_2\dirprod\sigma_2 + \sigma_3\dirprod\sigma_3)/4
\end{eqnarray*}
Whence, for any $i\in\setof{1,2,3}$ we get
\[ Q_i \eqdf \bowtie{B_0}{B_0} + \bowtie{B_i}{B_i} = \frac{I +
\gamma_i(\sigma_i \dirprod \sigma_i)}{2}, \]
where $(\gamma_1,\gamma_2,\gamma_3) = (1,-1,1)$.

Each pair $(Q_i, I - Q_i)$ is a binary projective measurement which we
can express in $\oplus$-separate form ($\oplus$ is the parity
function) as follows: Let $P^A = (I +
\vec{\alpha}\cdot\vec{\sigma})/2$ and $P^B = (I +
\vec{\beta}\cdot\vec{\sigma})/2$ be arbitrary one-qubit projectors
with unit trace ($\vec{\alpha} = (\alpha_1,\alpha_2,\alpha_3)$ and
$\vec{\beta} = (\beta_1,\beta_2,\beta_3)$ are arbitrary unit vectors
in $\reals^3$).  Simplifying the equation
\begin{equation}\label{f-sep}
P^A\dirprod P^B+(I-P^A)\dirprod(I-P^B) = Q_i = \frac{I +
\gamma_i(\sigma_i\dirprod\sigma_i)}{2}
\end{equation}
yields the equivalent equation
\begin{equation}\label{f-sep-equiv}
\vec{\alpha}\cdot\vec{\sigma} \dirprod \vec{\beta}\cdot\vec{\sigma}
= \gamma_i(\sigma_i \dirprod \sigma_i).
\end{equation}
By the linear independence of the $\sigma_j$, there are only two
possible solutions for $\vec{\alpha}$ and $\vec{\beta}$, namely,
$\alpha_j = \beta_j = 0$ for $j\neq i$ and $(\alpha_i,\beta_i)$ equals
either $(\gamma_i,1)$ or $(1,\gamma_i)$.  We arbitrarily choose the
latter of these.  (For $i\in\setof{1,3}$ these two solutions are the
same.)  Thus we have
\begin{eqnarray}
(Q_1,I-Q_1) & = & \septwo{\frac{I + \sigma_1}{2}}{\frac{I -
\sigma_1}{2}}{\frac{I+\sigma_1}{2}}{\frac{I-\sigma_1}{2}} \label{Q1} \\
(Q_2,I-Q_2) & = & \septwo{\frac{I + \sigma_2}{2}}{\frac{I -
\sigma_2}{2}}{\frac{I-\sigma_2}{2}}{\frac{I+\sigma_2}{2}} \label{Q2} \\
(Q_3,I-Q_3) & = & \septwo{\frac{I + \sigma_3}{2}}{\frac{I -
\sigma_3}{2}}{\frac{I+\sigma_3}{2}}{\frac{I-\sigma_3}{2}}. \label{Q3}
\end{eqnarray}
Applying any two of these measurements in sequence is equivalent to a
Bell measurement.

For the case of a general one-qubit gate $U$, we define
\[ R_i \eqdf \bowtie{U_0}{U_0} + \bowtie{U_i}{U_i} = (I \dirprod
U)Q_i(I \dirprod \adj{U}) \]
for $i\in\setof{1,2,3}$.  Similarly with Equation~\ref{f-sep}, we now
solve
\begin{equation}
P^A\dirprod P^B+(I-P^A)\dirprod(I-P^B) = R_i = \frac{I +
\gamma_i(\sigma_i\dirprod U\sigma_i\adj{U})}{2},
\end{equation}
but here we express $P^B$ not as before but instead as $(I +
U(\vec{\beta}\cdot\vec{\sigma})\adj{U})/2$, and we get the exact same
conditions on $\vec{\alpha}$ and $\vec{\beta}$ as in
Equation~\ref{f-sep-equiv}.  Thus,
\begin{eqnarray*}
(R_1,I-R_1) & = & \septwo{\frac{I + \sigma_1}{2}}{\frac{I -
\sigma_1}{2}}{\frac{I+U\sigma_1\adj{U}}{2}}{\frac{I-U\sigma_1\adj{U}}{2}} \\
(R_2,I-R_2) & = & \septwo{\frac{I + \sigma_2}{2}}{\frac{I -
\sigma_2}{2}}{\frac{I-U\sigma_2\adj{U}}{2}}{\frac{I+U\sigma_2\adj{U}}{2}} \\
(R_3,I-R_3) & = & \septwo{\frac{I + \sigma_3}{2}}{\frac{I -
\sigma_3}{2}}{\frac{I+U\sigma_3\adj{U}}{2}}{\frac{I-U\sigma_3\adj{U}}{2}}.
\end{eqnarray*}
Applying any two of these measurements in sequence is equivalent to a
measurement in the $\setof{\ket{U_j}}$-basis.  We have shown the
following:

\begin{theorem}\label{thm:one-qubit}
For any one-qubit unitary operator $U$, the projective measurement in
the $\setof{\ket{U_j}}$-basis of Equation~\ref{Uj} is equivalent to
the composition of the two $\oplus$-separate measurements
$$\septwo{\frac{I + \sigma_1}{2}}{\frac{I - \sigma_1}{2}}{\frac{I +
U\sigma_1\adj{U}}{2}}{\frac{I-U\sigma_1\adj{U}}{2}}$$ and 
$$\septwo{\frac{I + \sigma_3}{2}}{\frac{I -
\sigma_3}{2}}{\frac{I + U\sigma_3\adj{U}}{2}}{\frac{I -
U\sigma_3\adj{U}}{2}}.$$
\end{theorem}

Theorem~\ref{thm:one-qubit} also applies to the Bell measurement,
where $U = I$.  Another special case is when $U$ is one of the Pauli
matrices $\sigma_i$ for $i\in\setof{1,2,3}$.  For $j\in\setof{1,3}$,
we have $U\sigma_j\adj{U} = \sigma_i\sigma_j\sigma_i =
(-1)^{\delta_{ij}}\sigma_j$.  This means that we are still doing
essentially a Bell measurement, but we need to negate one or both of
the classical bits that are input to the $\oplus$ function (see
Section~\ref{sec:summary}).  From now on, we will refer to these
measurements also as Bell measurements.

We now consider what happens in Nielsen's protocol when we try to
simulate $U$ over several trials.  For
$j,k,\ell,\ldots\in\setof{1,2,3}$, define the (unitary and Hermitian)
operators
\begin{eqnarray*}
U_j & = & U\sigma_j\adj{U} \\
U_{j,k} & = & U_j\sigma_k\adj{U_j} \\
U_{j,k,\ell} & = & U_{j,k}\sigma_{\ell}\adj{U_{j,k}} \\
& \vdots &
\end{eqnarray*}
(Note that $U_j$ and $\ket{U_j}$ mean entirely different things.)
Suppose for the first trial we prepare state $\ket{U_j}$ for some
$j\in\four$, and our Bell measurement yields some $m\neq j$.  Then on
the second trial we must simulate $U\sigma_m\sigma_j\adj{U} \prop U_{[m,j]}$
(cf.\ Equation~\ref{next-U}).  Suppose for the second trial we prepare
some $\ket{U_{j'}}$, and our Bell measurement yields some $m'\neq
j'$.  Then on the third trial we must simulate $U_{[m,j],[m',j']}$,
and so on.

To simulate an arbitrary $U$ there are potentially infinitely many
gates $U_{j,k,\ell,\ldots}$ that we may need to try in order to
succeed reliably.  (For example, if $U = e^{i\theta\sigma_3}$ where
$\theta$ is an irrational multiple of $\pi$, then
$U_{2},U_{2,2},U_{2,2,2},\ldots$ are all distinct.)  Fortunately, for
the two gates $H$ and $T$ in our universal set, this is not the case.
For the Hadamard gate, we have
\begin{eqnarray*}
H_1 = H\sigma_1H & = & \sigma_3 \\
H_2 = H\sigma_2H & = & -\sigma_2 \\
H_3 = H\sigma_3H & = & \sigma_1
\end{eqnarray*}
so if we don't succeed in the first trial, all subsequent trials will
consist \emph{entirely} of Bell measurements (see the remark following
Theorem~\ref{thm:one-qubit}).  For the $T$ gate, we have
\begin{eqnarray*}
T_1 & = & \left[\begin{matrix}
0	&	e^{-i\pi/4}	\\
e^{i\pi4}	&	0	\\
\end{matrix}\right] \\
T_2 & = & \left[\begin{matrix}
0	&	-ie^{-i\pi/4}	\\
ie^{i\pi/4}	&	0	\\
\end{matrix}\right] \\
T_3 & = & \sigma_3,
\end{eqnarray*}
and 
\begin{eqnarray*}
T_{1,1} & = & \sigma_2  \\
T_{1,2} & = & \sigma_1  \\
T_{1,3} & = & -\sigma_3  \\
T_{2,1} & = & -\sigma_2 \\
T_{2,2} & = & -\sigma_1 \\
T_{2,3} & = & -\sigma_3,
\end{eqnarray*}
so if we don't succeed in the first two trials, all subsequent trials
will use just Bell measurements.

\subsection{Simulating the C-NOT Gate}

Let $U$ be an arbitrary $2$-qubit unitary operator, and let
$\ket{U_{jk}}$ be as in Equation~\ref{Ujk}.  We have
\[ \bowtie{{U_{jk}}}{{U_{jk}}} = (I_{12}\dirprod U_{34})(\lbt{B_j}{13}
\dirprod \lbt{B_k}{24}) (I_{12}\dirprod
(\adj{U})_{34}). \]
First we decompose the $\ket{U_{jk}}$ measurement into an equivalent
series of four binary measurements by adding up the projectors above
in various ways.  For $j\in\four$ let
\begin{eqnarray}
Q_j &=& \sum_{k=0}^3 \bowtie{{U_{jk}}}{{U_{jk}}} \\
&=& \sum_{k=0}^3 (I_{12}\dirprod U_{34})(\lbt{B_j}{13} \dirprod
	\lbt{B_k}{24})(I_{12}\dirprod \adj{U}_{34}) \\
&=& (I_{12}\dirprod U_{34})(\lbt{B_j}{13} \dirprod I_{24})
	(I_{12}\dirprod {\adj{U}}_{34}), \label{first-Qj}
\end{eqnarray}
and for $k\in\four$ let
\begin{eqnarray}
R_k &=& \sum_{j=0}^3 \bowtie{{U_{jk}}}{{U_{jk}}} \\
& = & (I_{12}\dirprod U_{34})(I_{13} \dirprod \lbt{B_k}{24} )
(I_{12}\dirprod {\adj{U}}_{34}) \\
& = & I_1 \dirprod (I_2 \dirprod U_{34})(I_3 \dirprod
\lbt{B_k}{24})(I_2\dirprod {\adj{U}}_{34}) \label{Rk}
\end{eqnarray}
We see that the measurement $(R_0,R_1,R_2,R_3)$ only involves qubits
$2$, $3$, and $4$, and leaves qubit~$1$ alone.  Composing this
measurement with $(Q_0,Q_1,Q_2,Q_3)$ above is equivalent to the
complete (nondegenerate) measurement in the
$\setof{\ket{U_{jk}}}$-basis.

Now let $U$ be the C-NOT gate, i.e., $U\ket{ab} = \ket{a}\ket{a\oplus
b}$ for $a,b\in\two$.  Evidently,
\begin{equation}\label{cnot}
U = \adj{U} = \sum_{b=0}^1\bowtie{b}{b} \dirprod \sigma_b.
\end{equation}
Substituting Equation~\ref{cnot} into Equation~\ref{first-Qj} gives,
after some calculation,
\[ Q_j = I_2 \dirprod \sum_{b=0}^1\sum_{c=0}^1 (\sigma_{[b,c]})_4
\dirprod \left[ (I_1 \dirprod
\lbt{b}{3})\lbt{B_j}{13}(I_1 \dirprod \lbt{c}{3})\right], \]
and so, after more calculation,
\begin{eqnarray}
Q_0 + Q_1 & = & I_2 \dirprod (I_{134} + (\sigma_1 \dirprod \sigma_1
\dirprod \sigma_1)_{134})/2, \label{Q-01} \\
Q_0 + Q_3 & = & I_{24} \dirprod (\lbt{B_0}{13} + \lbt{B_3}{13}).
\label{Q-03}
\end{eqnarray}
Equation~\ref{Q-03} describes the second Bell measurement ($U=I$) of
Theorem~\ref{thm:one-qubit} on qubits $1$ and $3$ (qubits $2$ and $4$
are left alone).  Equation~\ref{Q-01} describes a measurement on
qubits $1$, $3$, and $4$ that can be expressed in $\oplus$-separate
form as
\begin{equation}\label{Q-01-separate}
\septhree{\frac{I + \sigma_1}{2}}{\frac{I -
\sigma_1}{2}}{\frac{I + \sigma_1}{2}}{\frac{I - \sigma_1}{2}}{\frac{I
+ \sigma_1}{2}}{\frac{I-\sigma_1}{2}},
\end{equation}
which is clearly symmetric under permutation of qubits.

Adding the $R_k$ in two different pairs gives our other two binary
measurements.
Ignoring qubit~$1$, Equation~\ref{Rk} becomes
\begin{eqnarray*}
\lefteqn{\left(I_2 \dirprod \sum_{b=0}^1
\lbt{b}{3} \dirprod
\left(\sigma_b\right)_4\right)\left(\sum_{c=0}^1 \lbt{c}{3}
\dirprod \lbt{B_k}{24}\right)\left(I_2\dirprod \sum_{d=0}^1
\lbt{d}{3} \dirprod \left(\sigma_d\right)_4\right)} \\
& = & \sum_{b,c,d} \left(I_2 \dirprod \lbt{b}{3} \dirprod
\left(\sigma_b\right)_4\right)\left(\lbt{c}{3} \dirprod
\lbt{B_k}{24}\right)\left(I_2\dirprod
\lbt{d}{3} \dirprod \left(\sigma_d\right)_4\right) \\
& = & \sum_c \lbt{c}{3} \dirprod \left[ \left(I_2 \dirprod
\left(\sigma_c\right)_4\right) \lbt{B_k}{24} \left(I_2
\dirprod \left(\sigma_c\right)_4\right) \right] \\
& = & \sum_c \lbt{c}{3} \dirprod
\lbt{B_{[c,k]}}{24}.
\end{eqnarray*}
So,
\begin{eqnarray*}
R_0+R_1 & = & I_1 \dirprod \left[ \lbt{0}{3} \dirprod
\left(\lbt{B_0}{24} + \lbt{B_1}{24}\right) +
\lbt{1}{3} \dirprod \left(\lbt{B_1}{24} + \lbt{B_0}{24}\right) \right] \\ 
& = &  I_{13} \dirprod \left(\lbt{B_0}{24} + \lbt{B_1}{24}\right),
\end{eqnarray*}
and thus $(R_0+R_1,R_2+R_3)$ is the measurement
\begin{equation}\label{R-01}
\septwo{\frac{I + \sigma_1}{2}}{\frac{I -
\sigma_1}{2}}{\frac{I+\sigma_1}{2}}{\frac{I-\sigma_1}{2}}
\end{equation}
of Equation~\ref{Q1} on qubits $2$
and $4$.  In the same way, we get 
\[ R_0+R_3 = I_1 \dirprod \left[ \lbt{0}{3} \dirprod
(\lbt{B_0}{24} + \lbt{B_3}{24}) +
\lbt{1}{3} \dirprod (\lbt{B_1}{24} + \lbt{B_2}{24}) \right], \]
and so, using Equation~\ref{Q3}, we see that $(R_0+R_3,R_1+R_2)$ is
the measurement
\begin{eqnarray}
\lefteqn{\septhree{\bowtie{0}{0}}{\bowtie{1}{1}}{\frac{I +
\sigma_3}{2}}{\frac{I-\sigma_3}{2}}{\frac{I + \sigma_3}{2}}{\frac{I -
\sigma_3}{2}}} \\
& = & \septhree{\bowtie{0}{0}}{\bowtie{1}{1}}{\bowtie{0}{0}}%
{\bowtie{1}{1}}{\bowtie{0}{0}}{\bowtie{1}{1}} \\
& = & \septhree{\frac{I+\sigma_3}{2}}{\frac{I - \sigma_3}{2}}{\frac{I
+ \sigma_3}{2}}{\frac{I - \sigma_3}{2}}{\frac{I +
\sigma_3}{2}}{\frac{I - \sigma_3}{2}}, \label{R-03}
\end{eqnarray}
which is just the parity of qubits $2$, $3$, and $4$.

Just as in the one-qubit case, if one attempt to simulate $U$ fails,
we need to simulate a gate of the form $U(\sigma_{[j,m]} \dirprod
\sigma_{[k,n]})\adj{U}$ on the next attempt, for some $(m,n)\neq
(j,k)$.  In the case of C-NOT, we will always get back to simulating
the tensor product of Pauli matrices:
\begin{eqnarray*}
U(\sigma_0\dirprod\sigma_0)U &=& I\dirprod I \\
U(\sigma_0\dirprod\sigma_1)U &=& I\dirprod\sigma_1 \\
U(\sigma_0\dirprod\sigma_2)U &=& \sigma_3\dirprod\sigma_2 \\
U(\sigma_0\dirprod\sigma_3)U &=& \sigma_3\dirprod\sigma_3 \\
U(\sigma_1\dirprod\sigma_0)U &=& \sigma_1\dirprod\sigma_1 \\
U(\sigma_1\dirprod\sigma_1)U &=& \sigma_1\dirprod I \\
U(\sigma_1\dirprod\sigma_2)U &=& \sigma_2\dirprod\sigma_3 \\
U(\sigma_1\dirprod\sigma_3)U &=& -\sigma_2\dirprod\sigma_2 \\
U(\sigma_2\dirprod\sigma_0)U &=& \sigma_2\dirprod\sigma_1 \\
U(\sigma_2\dirprod\sigma_1)U &=& \sigma_2\dirprod I \\
U(\sigma_2\dirprod\sigma_2)U &=& -\sigma_1\dirprod\sigma_3 \\
U(\sigma_2\dirprod\sigma_3)U &=& \sigma_1\dirprod\sigma_2 \\
U(\sigma_3\dirprod\sigma_0)U &=& \sigma_3\dirprod I \\
U(\sigma_3\dirprod\sigma_1)U &=& \sigma_3\dirprod\sigma_1 \\
U(\sigma_3\dirprod\sigma_2)U &=& I\dirprod\sigma_2 \\
U(\sigma_3\dirprod\sigma_3)U &=& I\dirprod\sigma_3.
\end{eqnarray*}
Therefore, only Bell measurements will be needed after the first
attempt.

\section{Summary of Measurements}\label{sec:summary}

In this section we review the collection of measurements that we have
shown to be universal for quantum computation.  For our pseudoseparate
measurements, we have used only four single-qubit measurements in
various combinations:
\begin{eqnarray}
X_0 & \eqdf & (I + \sigma_1)/2 = H\bowtie{0}{0}H = \frac{1}{2} \left[
\begin{matrix} 1 & 1 \\ 1 & 1 \end{matrix} \right] \\
Y_0 & \eqdf & (I + \sigma_2)/2 = \frac{1}{2} \left[
\begin{array}{rr} 1 & -i \\ i & 1 \end{array} \right] \\
Z_0 & \eqdf & (I + \sigma_3)/2 = \bowtie{0}{0} = \left[
\begin{matrix} 1 & 0 \\ 0 & 0 \end{matrix} \right] \\
W_0 & \eqdf & (I + T_1)/2 = \frac{1}{2}\left(I + \frac{\sigma_1 +
\sigma_2}{\sqrt{2}}\right) = \frac{1}{2} \left[
\begin{matrix} 1 & e^{-i\pi/4} \\ e^{i\pi/4} & 1 \end{matrix} \right].
\end{eqnarray}
Define $X_1$, $Y_1$, $Z_1$, and $W_1$ to be $I-X_0$, $I-Y_0$, $I-Z_0$,
and $I-W_0$, respectively.

\paragraph{Bell Measurements.}{A Bell measurement is equivalent to
\begin{equation}\label{Bell-measurement}
\septwo{X_0}{X_1}{X_0}{X_1} \mbox{ and }
\septwo{Z_0}{Z_1}{Z_0}{Z_1}
\end{equation}
applied in succession (cf.\ Theorem~\ref{thm:one-qubit}).}

\paragraph{Simulating Pauli Matrices.}{Off-line preparation for the
first trial simulating a Pauli matrix requires
\begin{eqnarray}
\septwo{X_0}{X_1}{X_0}{X_1} \mbox{ and }
\septwo{Z_0}{Z_1}{Z_1}{Z_0} & \mbox{for} & \sigma_1, \\
\septwo{X_0}{X_1}{X_1}{X_0} \mbox{ and }
\septwo{Z_0}{Z_1}{Z_1}{Z_0} & \mbox{for} & \sigma_2, \\
\septwo{X_0}{X_1}{X_1}{X_0} \mbox{ and }
\septwo{Z_0}{Z_1}{Z_0}{Z_1} & \mbox{for} & \sigma_3.
\end{eqnarray}
By the remark following Theorem~\ref{thm:one-qubit}, we refer to these
also as Bell measurements.  Any subsequent trial needed to simulate
$\sigma_i$ then requires only Bell measurements.}

\paragraph{Simulating $H$.}{Off-line preparation for the first trial
simulating $H$ requires
\begin{equation}
\septwo{X_0}{X_1}{Z_0}{Z_1} \mbox{ and }
\septwo{Z_0}{Z_1}{X_0}{X_1},
\end{equation}
applied in succession, again by Theorem~\ref{thm:one-qubit}.
Subsequent trials need only Bell measurements.}

\paragraph{Simulating $T$.}{Off-line preparation for the first trial
simulating $T$ requires
\begin{equation}
\septwo{X_0}{X_1}{W_0}{W_1} \mbox{ and }
\septwo{Z_0}{Z_1}{Z_0}{Z_1},
\end{equation}
by Theorem~\ref{thm:one-qubit}, using the fact that $T\sigma_3\adj{T}
= T_3 = \sigma_3$.  The second trial simulating $T$ must simulate
either $T_1$, $T_2$, or $T_3$.  Simulating $T_1$ requires
\begin{equation}
\septwo{X_0}{X_1}{Y_0}{Y_1} \mbox{ and }
\septwo{Z_0}{Z_1}{Z_1}{Z_0},
\end{equation}
since $T_1\sigma_1\adj{T_1} = T_{1,1} = \sigma_2$ and $T_1\sigma_3
\adj{T_1} = T_{1,3} = -\sigma_3$.  Simulating $T_2$ requires
\begin{equation}
\septwo{X_0}{X_1}{Y_1}{Y_0} \mbox{ and }
\septwo{Z_0}{Z_1}{Z_1}{Z_0}.
\end{equation}
Simulating $T_3$ and any subsequent trials require only Bell
measurements.}

\paragraph{Simulating C-NOT.}{Off-line preparation for the first trial
simulating C-NOT requires the measurement
\begin{equation}
\left(\frac{I + \sigma_1\dirprod\sigma_1\dirprod\sigma_1}{2},\; \frac{I -
\sigma_1\dirprod\sigma_1\dirprod\sigma_1}{2}\right) =
\septhree{X_0}{X_1}{X_0}{X_1}{X_0}{X_1}
\end{equation}
on qubits $1$, $3$, and $4$ (Equations~\ref{Q-01} and
\ref{Q-01-separate}) and a Bell measurement on ancilla qubits $1$ and
$3$ (Equation~\ref{Q-03}), together with
\begin{equation}
\septwo{X_0}{X_1}{X_0}{X_1}
\end{equation}
on qubits $2$ and $4$ (Equation~\ref{R-01}), and
\begin{equation}
\septhree{Z_0}{Z_1}{Z_0}{Z_1}{Z_0}{Z_1}
\end{equation}
on qubits $2$, $3$, and $4$ (Equation~\ref{R-03}).  All other
measurements are Bell measurements.}

\section{Further Work}

We have shown that three-qubit, $\oplus$-separate measurements suffice
for universal quantum computation.  Do two-qubit measurements suffice?

We have also shown that two-qubit pseudoseparate measurements (to
simulate arbitrary one-qubit gates), together with a fixed finite
set of two- and three-qubit pseudoseparate measurements (to implement
C-NOT), suffice to \emph{exactly} simulate any unitary operator $A$ on
$n$-qubits (see \cite{NC:quantumbook} for example).  However, for a
given $A$, it may be the case that an infinite family of measurements
are needed to simulate $A$ exactly with probability 1, using Nielsen's
scheme.  Perhaps there is an alternate scheme whereby for every $A$
there is a fixed finite set of measurements sufficient to simulate $A$
exactly with probability 1.

Another interesting avenue of research is to see whether various
quantum algorithms can be made to tolerate errors in their gates of
the form of a failed single trial of Nielsen's protocol.  For example,
a one-qubit $U$-gate may actually apply $U\sigma_j$ for some
(classically known) $j\in\four$ with uniform probability.  Perhaps
useful computations can be done despite this nondeterminism, in which
case, we would not need to have repeated trials of Nielsen's protocol
when implementing such algorithms.

\section*{Acknowledgments}

We would like to thank Alonso Botero, Andrei Stoica, and Shengjun Wu
for several interesting and stimulating discussions.

\newcommand{\etalchar}[1]{$^{#1}$}

\end{document}